**Energy-efficient generation of skyrmion phases in Co/Ni/Pt-based multilayers using Joule heating**

Jeffrey A. Brock[1], Sergio A. Montoya[2], Mi-Young Im[3], and Eric E. Fullerton[1]

[1] Center for Memory and Recording Research, University of California San Diego, La Jolla, CA USA

[2] Naval Information Warfare Center – Pacific, San Diego, CA USA

[3] Center for X-Ray Optics, Lawrence Berkeley National Laboratory, Berkeley, CA, USA

**Abstract:**

We have studied the effects of electrical current pulses on skyrmion formation in a series of Co/Ni/Pt-based multilayers. Transmission X-ray microscopy reveals that by applying electrical current pulses of duration and current density on the order of $\tau$=50 µs and $j$=1.7 x $10^{10}$ A/m$^2$, respectively, in an applied magnetic field of $\mu_0 H_z$=50 mT, stripe-to-skyrmion transformations are attained. The skyrmions then remain stable across a wide range of magnetic fields, including zero field. We primarily attribute the transformation to current-induced Joule heating on the order of ~128 K. Reducing the magnetic moment and perpendicular anisotropy using thin rare-earth spacers reduces the pulse duration, current density, and magnetic field necessary to 25 µs, 2.4 x $10^9$ A/m$^2$, and 27 mT, respectively. These findings show that energetic inputs allow for the formation of skyrmion phases in a broad class of materials and that material properties can be tuned to yield more energy-efficient access to skyrmion phases.

**Introduction:**

Over the past decade, there has been increasing study of the magnetic textures known as skyrmions due to their quantized spin topology.[1,2,3] This spin structure gives rise to topologically enhanced stability and unique dynamics that are attractive for emerging data storage, memory, and computational applications.[4] Creating a material system capable of hosting such a unique magnetic texture involves a delicate interplay between the exchange, dipolar, and anisotropy energies of the material. While it is possible to optimize these three energies alone to create a material capable of hosting skyrmions[5], often an additional energetic contribution from the Dzyaloshinskii-Moriya interaction (DMI)[6,7] is key to fostering skyrmions. In contrast to ferromagnetic exchange, the DMI favors a 90° canting of neighboring spins relative to one another.[8] Initially, the skyrmion texture was first reported in a bulk crystal of the weak ferromagnet MnSi, where the presence of DMI results from the material's non-centrosymmetric crystal structure.[9] Over the ensuing years, the interfacial DMI due to symmetry breaking at ferromagnet/heavy metal or oxide interfaces in structurally-asymmetric structures[10,11,12,13,14] has been exploited to foster skyrmions in many multilayer systems (e.g., Refs 11, 15, 16, 17, 18, 19, 20, and references therein).

Recently, there has been interest in leveraging the effects of rapid heating to transform magnetic stripe domains into skyrmions.[21,22,23,24] Micromagnetic and atomistic simulations have shown that a short timescale temperature rise in these systems can transform the sample between a variety of morphological magnetic domain phases, including the skyrmion state.[25] Crucially, experimental reports have demonstrated that skyrmions formed via Joule heating exhibit several attractive traits, such as stability against changes in the magnetic field (including at zero field), and even the ability to nucleate skyrmions in zero applied magnetic field in some systems and temperature ranges.[21] However, there are still many open questions as to the physical nature of this thermally driven transformation, including approaches by which the magnetic field and Joule heating necessary to drive the stripe-to-skyrmion transformation can be reduced while still retaining the robust stability of the skyrmion phase.

We have chosen to study a multilayer of the base structure [Co (0.7)/ Ni (0.5) /Pt (0.7)] (thicknesses in nm). Our choice of materials was motivated by reports suggesting that the Pt/Co and Pt/Ni interfaces can have different strengths of interfacial DMI, such that an additive DMI is obtained when they are incorporated in structurally asymmetric materials stacks.[26] Additionally,

Co/Ni-based systems have already been the subject of extensive study on account of the inherent tunability of their basic physical properties, such as the saturation magnetization, spin polarization, strength of perpendicular magnetic anisotropy, and Gilbert damping.[27,28] Indeed, this ease with which the magnetic properties of Co/Ni-based systems can be adjusted has been capitalized to experimentally foster skyrmions in a variety of structures, including Pt/[Co/Ni]/Ir[29] and Pt/[Co/Ni/Co]/Au[30], albeit only stabilizing the skyrmion phase in an applied magnetic field.

Using resonant full-field magnetic transmission X-ray microscopy (MTXM) operating at the Co $L_3$ absorption edge (778 eV), we imaged the effects of applied magnetic fields and electrical current pulses on the magnetic domain morphology in several 10-μm-wide, 3-mm-long [Co/Ni/Pt]-type patterned devices patterned on SiN membranes. This experimental data, coupled with transport measurements and modeling of the sample's thermal properties, demonstrates that current pulses delivering Joule heating on the order of ~ 128 K within 50 μs in a 50 mT field are capable of completely transforming all stripes within the field of view of a [Co/Ni/Pt]$_{20}$ sample into skyrmions and that these skyrmions remain stable over an extended range of magnetic fields. By reducing the saturation magnetization and effective perpendicular anisotropy through the insertion of thin rare-earth spacer layers within the ferromagnetic layers, the Joule heating needed to accomplish the same transformation was reduced to ~15 K, while still retaining robust field stability. These results support the idea that Joule heating in a magnetic field allows for the efficient nucleation of skyrmions in a broad class of multilayers and identifies pathways towards lowering the current density needed to achieve skyrmions by such means.

**Results and Discussion:**

Figures 1(a-c) show MTXM images of the field-dependent magnetic domain morphology of a [Co (0.7 nm)/Ni (0.5 nm)/Pt (0.7nm)]$_{20}$ (henceforth referred to as [Co/Ni/Pt]$_{20}$) patterned wire as a perpendicular magnetic field ($\mu_0 H_z$) was swept from negative saturation to positive saturation at room temperature. The contrast in these images reflects the perpendicular magnetization averaged over the thickness of the film, with opposite magnetization textures are represented as gray (+$M_z$) and black (-$M_z$) contrast. At remanence ($\mu_0 H_z = 0$ mT), the domain morphology primarily consists of disordered stripe domains with a width of 105 nm ± 12 nm [Fig. 1(a)]. As the perpendicular field is increased ($\mu_0 H_z = +90, +250$ mT), the disordered stripe domains begin to contract [Fig. 1(b)] until only a few stripes are observable in the field of view [Fig. 1(c)]. Throughout the field-induced evolution of the stripe domains, the [Co/Ni/Pt]$_{20}$ patterned wire does not exhibit a skyrmion phase. Supplementary Figure S1[31] shows the field-dependent domain morphology in greater detail, obtained using a [Co/ Ni/ Pt]$_{20}$ continuous film.

Having established that the [Co/Ni/Pt]$_{20}$ patterned wire does not exhibit a skyrmion phase solely in response to an applied magnetic field at room temperature, we proceeded to investigate the effects of applying electrical current pulses of varying electrical current density $j$, and pulse duration $\tau$ on the field-dependent domain morphology. At any given field, the domain morphology was recorded before and after injecting a current pulse. After each trial, the domain morphology was reset by cycling the perpendicular field up to positive saturation, then reducing the field to negative saturation, and finally increasing field to the desired positive value. Figures 1(d-f) show the domain morphology that was obtained after applying a single $\tau = 50$ µs long current pulse with current density $j = 1.7 \times 10^{10}$ A/m$^2$ onto the domain states depicted in Figs. 1(a-c). In calculating $j$, we considered the thickness of the entire sample (including the seeding and capping layers). At remanence ($\mu_0 H_z = 0$ mT), we find the electrical pulse primarily results in the local rearrangement of disordered stripe domains, without the formation of any new types of magnetic features [Figs. 1(a, d)]. Under moderate fields ($\mu_0 H_z = +90$ mT), the current excitation transforms a disordered stripe phase into a dense, disordered skyrmion phase [Figs. 1(b, e)]; from additional experiments, we estimate that the temperature increase associated with this pulse is ~128 K (see Methods, Supp. Fig. S2[31]). Like most multilayered materials that host field-induced[32,33] and current-induced[34] skyrmions, we observe that the individual features exhibit a shape that is not perfectly circular, resulting from pinning centers present throughout the film. The skyrmion features exhibit an average diameter of 69 nm ± 11 nm, which is comparable to the average stripe width of 66 nm ± 8 nm before the application of the current pulse. Near magnetic saturation ($\mu_0 H_z = +250$ mT), we find the remaining disordered stripe domains are annihilated after applying the current pulse, with the patterned wire becoming uniformly magnetized in the direction of the applied field [Figs. 1(c, f)].

Next, the field-stability of the skyrmion phase in the [Co/Ni/Pt]$_{20}$ patterned wire was explored as the magnetic field was reduced from the positive field at which they were elicited using electrical currents towards negative saturation. Figures 2(a-c) show the evolution of the skyrmion phase as the applied magnetic field was varied. As the field was reduced, the individual skyrmion features grow nonuniformly within their local vicinity while retaining their overall general shape; moreover, the skyrmion features do not elongate/recombine to form a disordered stripe phase, as is commonly observed in multilayered materials hosting field-stabilized skyrmions. At remanence, the domain morphology consists of a close-packed skyrmion phase [Fig. 2(b)], which differs from the ground state that is obtained under perpendicular fields [Fig. 1(a)]. Reducing the magnetic field further ($\mu_0 H_z$ = -70 mT), we observe that the close-packed skyrmion phase continues to persist [Fig. 2(c)].

We then explored, in detail, the morphological transformations that arise in the [Co/Ni/Pt]$_{20}$ patterned wire as a function of perpendicular field and current pulse width. Figure 3(a) shows the different degrees of stripe-to-skyrmion transformation that can be accessed under different $\tau$ - $\mu_0 H_z$ combinations (including the behaviors noted in Fig. 1) when applying a single electrical pulse with fixed density $j = 1.7 \times 10^{10}$ A/m$^2$. In Fig. 3, we also display the estimated temperature increase due to Joule heating $\Delta T_{est}$ for each pulse duration, determined experimentally using the techniques described in Methods and Supp. Fig. S2. The minimum current density necessary to induce the stripe-to-skyrmion transformation was determined as follows: First, the applied field was set to the point at which the hysteretic region of the magnetization curve for the sample in question first begins to open up when sweeping the field towards positive saturation. Next, the current pulse duration was fixed to 50 µs while applying electrical pulses of gradually increasing magnitude until a complete stripe-to-skyrmion transformation was observed. From this point, the pulse duration and applied magnetic fields were systematically varied to explore the responses. Like before, the field was cycled after each trial.

At remanence, we find that a single current pulse with durations varying from $\tau$ = 10 µs up to 100 µs is unable to induce a stripe-to-skyrmion transformation within our field of view. A similar observation is noted when applying a $\tau$ = 10 µs current pulse as a function of increasing field ($\mu_0 H_z$ = 0 mT to +120 mT). Under moderate fields ($\mu_0 H_z$ = +50 mT and +90 mT) we can access a $\tau$ - $\mu_0 H_z$ phase region where the electrical pulse first transforms a disordered stripe phase into a coexisting stripe and skyrmion phase; then, as the current pulse width is increased, a high-density skyrmion phase is formed. In general, the density of the skyrmion phase appears to depend on the density of stripe domains present before the current pulse is applied (see Supp. Fig. S3[31]). Under higher applied fields ($\mu_0 H_z$ = +120 mT and +250 mT), two additional $\tau$ - $\mu_0 H_z$ phase regions can be accessed – the current-induced contraction of stripe domains as well as a sparse stripe-to-sparse skyrmion transformation. Near the saturation field ($\mu_0 H_z = +300$ mT), the application of electrical current pulses annihilates the low-density stripe domains. Overall, the phase diagram illustrates the array of magnetic phases that can be accessed using electrical pulses. The physical mechanisms involved in the stripe-to-skyrmion transformation brought about by electrical current pulses will be discussed subsequently.

To explore how material properties can be tuned to facilitate stripe-to-skyrmion transformations using lower current density electrical pulses and lower perpendicular fields, we investigated different multilayer structures with moderate DMI. In general, it has been shown that reducing the saturation magnetization ($M_S$) and effective perpendicular anisotropy ($K_{eff}$) can lower the energy barrier that separates the stripe and skyrmion ground states in multilayered materials.[35,36] One potential pathway to lowering these two properties is through the incorporation of rare-earth spacer layers between the Co and Ni layers. Rare earth layers are expected to couple antiferromagnetically with the Co and Ni to reduce $M_S$ (through the creation of a net ferrimagnetic structure) while at the same time increasing the thickness of magnetic material in the structure and removing the Co/Ni interfaces, reducing $K_{eff}$.[37,38,39,40] Static characterization indicates that the [Co/Ni/Pt]$_{20}$ sample exhibits $M_S$ and $K_{eff}$ values of 1010 kA/m and 5.1 x 10$^5$ J/m$^3$, respectively (Supp. Fig. S4[31]). In line with the previous discussion, incorporating a thin spacer layer of rare earth material in a sample of the form [Co (0.7 nm)/Gd (0.4 nm)/Ni (0.5 nm)/Pt (0.7nm)]$_{20}$ reduces $M_S$ to 488 kA/m, while $K_{eff}$ was roughly halved to 2.5 x 10$^5$ J/m$^3$. The variation is considerably more dramatic when Tb was employed in a structure of the composition [Co (0.7 nm)/Tb (0.4 nm)/Ni (0.5 nm)/Pt (0.7nm)]$_{20}$, which exhibited $M_S$ = 512 kA/m and $K_{eff}$ = 8.2 x 10$^4$ J/m$^3$ – roughly an 80 % reduction in $K_{eff}$ relative to the [Co/Ni/Pt]$_{20}$ sample. All samples exhibit the "sheared" perpendicular hysteresis loops characteristic of a multi-domain state at zero field [Supp. Fig. S4(a)[31]]. Static characterization of $M_S$ as a function of temperature indicates that that the [Co/Tb/Ni/Pt]$_{20}$ sample does not experience a dramatic change in $M_s$ within the temperature range of 100 K – 400 K, implying that the composition is far from ferrimagnetic compensation [Supp. Fig. S5(a)[31]]. Furthermore, in-plane hysteresis loops of the [Co/Tb/Ni/Pt]$_{20}$ sample collected at higher temperatures suggest that the anisotropy does not remarkably change across the relevant temperature scale in the samples with rare earth spacer layers [Supp Fig. S5(b)[31]]. The field-dependent domain morphology of [Co/Tb/Ni/Pt]$_{20}$ and [Co/Gd/Ni/Pt]$_{20}$ patterned wires reveal that a purely field-stabilized skyrmion phase is not energetically accessible in these materials (Supp.

Fig. S6[31]). Similar to [Co/Ni/Pt]$_{20}$, both [Co/Tb/Ni/Pt]$_{20}$ and [Co/Gd/Ni/Pt]$_{20}$ exhibit disordered stripe phases at remanence, with stripes exhibiting an average width of 142 nm ± 18 nm and 137 nm ± 13 nm, respectively. Given the reduction in $M_S$, we expected the stripe domain width to broaden based on magnetostatic considerations.[41]

Figure 3(b) shows the morphological phases of a [Co/Tb/Ni/Pt]$_{20}$ patterned wire as a function of perpendicular field and current pulse-width. In general, comparable morphological phases are observed as in [Co/Ni/Pt]$_{20}$ [Fig. 3(a)] with two main distinctions: The diverse morphological phases can be accessed using a lower magnitude range of $\mu_0 H_z$ and a lower-magnitude current density pulse of $j = 2.4 \times 10^9$ A/m$^2$ (*i.e.*, one order of magnitude lower than for Co/Ni/Pt structures). Furthermore, the [Co/Tb/Ni/Pt]$_{20}$ sample exhibits a morphological phase that was unseen in [Co/Ni/Pt]$_{20}$ – the elongation of stripe domains collinear to $j$ (Supp. Fig. S7[31]). Similar behavior has been previously observed in [Pt/CoFeB/MgO]$_{20}$, and was attributed to a balance of ferromagnetic exchange, perpendicular anisotropy, and DMI energy in a high aspect ratio geometry of the narrow wire based on micromagnetic simulations.[21]

While the phase diagram shown in Fig. 3 demonstrates a diversity of morphological phases can be accessed in [Co/Ni/Pt]-based multilayers, it says nothing as to the physical mechanism responsible for the transformation. Because our multilayer samples are seeded on and capped with materials that are known to exhibit opposite signed spin-Hall angles[42,43], a net spin-orbit torque (SOT) may be acting on the system. Given the relative thickness of the seed and capping layers, we argue that these layers carry more current than any of the individual layers within the interior of the multilayer structure; thus, the capping and seed layers would be the most likely source of SOT. Furthermore, any SOT generated from the interior Pt spacer layers would be mostly negated, in that these layers would generate opposing signs of SOT at the top and bottom interfaces of the ferromagnetic layers. To investigate the possible effects of SOT, we varied the capping layer composition of the [Co/Ni/Pt]$_{20}$ sample from Ta to Pt to change the sign of the spin-Hall angle of the layer. It is expected that the use of a Pt capping layer should reduce the net SOT acting on the magnetic layers if a uniform Neel domain wall chirality is maintained through the thickness of the film. Alternatively, if the dipolar energy is sufficient to promote the formation of Néel caps[5,44], the use of a Pt capping layer would enhance the SOT. In Supp. Fig. S8[31], we demonstrate that there is an identical change in domain morphology in response to equivalent stimuli between samples with Ta and Pt capping layers – implying that SOT strength is not central to driving the stripe-to-skyrmion transformation in the [Co/Ni/Pt]-type samples.

We have performed additional experiments in order to discern the effects of Joule heating on the skyrmion nucleation observed in the [Co/Ni/Pt]-type samples, as proposed in Ref. 21. In the first set of tests, a composite pulse (consisting of a baseline and a spike component) was utilized to deliver the same net electron flow/SOT and Joule heating, while varying the maximum temperature attained by the sample; this was accomplished by modifying the relative delay of the spike pulse to the beginning, middle, and end of the baseline pulse [Fig. 4(a)]. The baseline pulse with no spike, of duration $\tau = 72$ µs and current density $j = 2.5 \times 10^9$ A/m$^2$, did not result in morphological transformations when applied on its own to a [Co (0.4 nm)/ Tb (0.4 nm)/ Co (0.4 nm)/ Ni (0.5 nm)/ Pt (0.7 nm)]$_{20}$ patterned wire under an applied field of $\mu_0 H_z = +173$ mT [Figs. 4(b, f)]. Like the other samples discussed thus far, this sample does not exhibit a purely field-stabilized skyrmion phase; the skyrmion phase is accessed by applying electrical current pulses within a $\tau$ - $\mu_0 H_z$ window unique to the sample. The spike pulse had $\tau = 10$ µs with current density $j = 5 \times 10^9$ A/m$^2$ [Fig. 4(a)]. For a spike pulse applied at the beginning of the baseline pulse [blue pulse, Fig. 4(a)], the field-stabilized stripe phase rearranged slightly [Fig. 4(c, g)]. As the relative delay of the spike pulse was increased to the middle and end of the baseline pulse [green and red pulses, Fig. 4(a)], the number of skyrmions present after the pulse increases [Figs. 4(d, h) and Figs. 4(e, i), respectively]. These tests suggest the time-dependent thermal profile from Joule heating plays an important role in the stripe-to-skyrmion generation in our [Co/Ni/Pt]-type samples. Specifically, as the delay of the spike pulse increases, the maximum temperature attained by the sample should increase as well – suggesting that temperature plays a strong role in promoting the observed morphological transformations. Using COMSOL Multiphysics, we have simulated the sample temperature change that occurs in response to the three composite pulse configurations discussed above, finding that applying the spike pulse at the end of the baseline leads to a final temperature ~ 30 K higher than when the spike is applied at the beginning of the baseline [see Methods and Supp. Fig. S9(b)[31]]. However, since it is known that Joule heating can enhance the effect of SOT acting on a ferromagnetic sample[45], this test does not allow for a complete disentanglement of the effects of Joule heating versus SOT.

In the second set of tests, the current density $j$ and pulse width $\tau$ were varied in such a way that the Joule heating previously identified as necessary for a complete stripe-to-skyrmion transformation was supplied at differing timescales. Once again, we have employed COMSOL Multiphysics to simulate the temperature changes resulting from these pulses [see Methods, Supp. Fig. S9(c)[31]]. Figure 5(a) schematically shows the different pulses that were injected to the [Co/Tb/Co/Ni/Pt]$_{20}$ patterned wire under an applied field of $\mu_0 H_z$ = +173 mT. A complete stripe-to-skyrmion transformation is observed using a 50-µs-long pulse with current density $j$ = 5 x 10$^9$ A/m$^2$ [Figs. 5(b, e)], estimated to increase the sample temperature by ~ 115 K. As these pulse characteristics correspond to the most rapid heating of the sample, this pulse should lead to the largest increase in temperature before the heat dissipates. A similarly-complete stripe-to-skyrmion transformation occurs in response to a 200-µs-long pulse with current density $j$ = 2.5 x 10$^9$ A/m$^2$ [Figs. 5(c, f)], implying that the temperature increase attained under these conditions (~ 85 K) is still enough to promote a complete morphological transformation. However, when the Joule heating is supplied using an 800 µs pulse of current density $j$ = 1.25 x 10$^9$ A/m$^2$ [Figs. 5(d,g)], significant dissipation is expected to occur during the heating process. Consequentially, the temperature change (~ 55 K) is not high enough to promote the nucleation of skyrmions in this sample.

To better understand the impacts of Joule heating on the stability of morphological phases in the [Co/Ni/Pt]$_{20}$ samples, we have applied the model of Ref. 36 to explore the size and stability of the isolated skyrmion bubble state formed in our samples. These calculations were performed using parameters extracted from the static characterization parameters stated in the Results section, save for the DMI energy density of D = 0.622 mJ/m$^2$ - which was determined by measuring the in-plane field-induced domain expansion asymmetry[46] in thinner samples that reverse via large bubble domains. In Fig. 6(a), the energy as a function of skyrmion bubble diameter is shown for several $\mu_0 H_z$ proximal to the minimum $\mu_0 H_z$ = +50 mT needed to elicit the skyrmion phase in the [Co/Ni/Pt]$_{20}$ sample. Besides predicting skyrmion diameters close to those experimentally observed, it is apparent that increasing the field even by 10-mT increments can substantially shallow the energy barriers separating different morphological states. Similar calculations for the [Co/Tb/Ni/Pt]$_{20}$ sample [Fig. 6(b)] illustrate that the reduced $M_S$ and $K_{eff}$ of this sample leads to shallower energy barriers between morphological states. Nonetheless, for all fields and samples considered, quasi-static variations in $\mu_0 H_z$ or temperature would likely be insufficient to overcome the potential well.[23]

To glean a picture of how the ease of skyrmion nucleation via Joule heating varies between the samples discussed above, we consider two parameters: The estimated change in sample temperature due to Joule heating necessary to induce the transformation $\Delta T_{est, st \rightarrow sk}$ (determined from the experiments shown in Supp. Fig. S2[31]), and the magnetic field in which the pulse is applied ($\mu_0 H_{z\ st \rightarrow sk}$). To simplify comparisons, we consider the pulse of minimum $\tau$ and $j$ needed to drive a complete stripe-to-skyrmion transformation in the lowest $\mu_0 H_z$ possible for a given sample at room temperature using the approach discussed previously. In Fig. 7(a), a comparison of $\Delta T_{est, st \rightarrow sk}$ for our samples is provided. The data for the [Co/Ni/Pt]$_{30}$ and [Co/Ni/Pt]$_{20}$ samples indicates that one way to reduce the heating needed to drive a complete stripe-to-skyrmion transformation is by reducing the total thickness of the sample by a third, which lowers $\Delta T_{est, st \rightarrow sk}$ by nearly one third. However, reducing $M_S$ and $K_{eff}$ has a more dramatic effect, lowering $\Delta T_{est, st \rightarrow sk}$ by upwards of 90% in [Co/Tb/Ni/Pt]$_{20}$ relative to [Co/Ni/Pt]$_{20}$. Compared to other multilayered systems in which skyrmion phases were induced via Joule heating, the systems containing rare earth spacers allow for a complete stripe-to-skyrmion transformation using a favorable amount of heating (*e.g.*, $\Delta T_{est, st \rightarrow sk} \approx 70$ K for [Pt/CoFeB/MgO]$_{20}$[21] versus ~15 K for [Co/Tb/Ni/Pt]$_{20}$). The plot of $\mu_0 H_{z\ st \rightarrow sk}$ for different sample compositions [Fig. 7(b)] shows a similar correlation to the thickness, $M_S$, and $K_{eff}$ of the samples. Moreover, the [Co/Gd/Ni/Pt]$_{20}$ and [Co/Tb/Ni/Pt]$_{20}$ samples exhibit $\mu_0 H_{z\ st \rightarrow sk}$ values comparable to other materials in which skyrmion phases formed using Joule heating were studied (*e.g.*, 10 mT for [Pt/CoFeB/MgO]$_{20}$[21] and 17 mT for [Pt/Co/Ir]$_{10}$[22])

**Conclusion**

We have systematically studied the effect of electrical current pulses on the generation of skyrmion phases in a variety of [Co/Ni/Pt]-based multilayers, finding that Joule heating is primarily responsible for the transformations in the domain morphology observed. Furthermore, we have shown that lowering the saturation magnetization and effective perpendicular magnetic anisotropy significantly reduces the Joule heating and magnetic field necessary to drive a complete stripe-to-skyrmion transformation while still retaining attractive magnetic properties, such as zero-field skyrmion stability. Besides illustrating pathways towards reducing the energetic input necessary to create skyrmion phases, these results further support the observations that Joule heating can enable the generation of skyrmions in a broad class of materials, over a wide range of environmental conditions, and in systems that do not require such a precise balance in energy contributions to be attained during the deposition process. Our results provide a pathway to tailor material properties to achieve energy-efficient skyrmion phases in multilayers using electrical current pulses.

**Methods:**

Samples of the composition (from the substrate to top layer) Ta (2 nm)/Pt (5 nm)/[Co (0.7 nm)/Ni (0.5 nm)/Pt (0.7nm)]$_{20}$/Ta (5 nm) were deposited at ambient temperature by DC magnetron sputtering. The chamber base pressure was less than 1 x 10$^{-8}$ Torr and the argon process gas pressure was 3 mTorr. For magnetic characterization, samples were grown on Si substrates with a 300-nm-thick thermal oxide coating. For full-field soft X-ray transmission microscopy experiments, samples were grown on 100-nm-thick Si$_3$N$_4$ membrane windows. Samples intended for electrical current studies were patterned into 10-μm-wide, 3-mm-long wires using metal lift-off ultra-violet photolithography. Magnetometry measurements were performed in both the out-of-plane and in-plane geometries using vibrating sample magnetometry (VSM) to determine the saturation magnetization ($M_S$) and effective perpendicular magnetic anisotropy ($K_{eff}$) from the saturation of the in-plane hard axis loop. In calculating volumetric parameters from the VSM data, only the thickness of the magnetic layers was considered.

The evolution of the magnetic domain morphology as a function of the out-of-plane field and in response to electrical current pulses was assessed using resonant full-field magnetic transmission X-ray microscopy (MTXM) at Beamline 6.1.2 of the Advanced Light Source at Lawrence Berkeley National Laboratory.[47] Due to the non-uniformity of the illuminating spot projected on the sample, only representative areas of the entire micrograph collected are shown in the figures. All MTXM data was collected using circularly polarized X-rays resonant to the Co L$_3$ adsorption edge (778 eV).

Electrical current pulses were applied using an Agilent Technologies 81150A Pulse Function Arbitrary Generator. For all current densities and pulse durations discussed in this study, the leading/trailing edges of the pulse were set to 6 ns. Before each new trial, the magnetization of the sample was saturated in a positive field, followed by saturation in a negative field, followed by stabilization at the target positive field. The shape of the applied current pulses were verified using an oscilloscope connected at the ground end of the circuit. The electrical resistance of the samples was monitored throughout the experiments to ensure that the electrical current pulses did not appreciably alter the material properties. The pulse energy was calculated from the sample resistance, and the duration and voltage of the electrical pulse supplied; calculation of the pulse energy density was performed by dividing this energy by the volume of the sample.

To understand the relationship between Joule heating and the temperature change of the samples, we have determined the temperature coefficient of resistance $\alpha$ of a $[Co/Ni/Pt]_{20}$ patterned device deposited on to a $Si_3N_4$ membrane by measuring the resistance as a function of temperature, employing a low-current AC lock-in technique [Supp. Fig. S2(a)[31]]. Subsequent measurements of the resistance of the same sample as a function of current density enabled an estimation of the temperature change as a function of current density [Supp. Fig. S2(b)[31]]. To account for the fact that the Joule heating for a particular current density varies with temperature due to the associated change in resistance of the sample, we have calculated the Joule heating energy densities using both the resistance of the sample at room temperature, as well as the resistance at the predicted increased temperature [Supp. Fig. S2(c)[31]]; in the manuscript and figures, we use the temperature values predicted by the former convention. All COMSOL Multiphysics simulations of temperature changes due to Joule heating used a simplified model of the multilayer as a Pt film with the same thickness and average resistivity as the actual multilayer; specifics on the geometry employed are shown in Supp. Fig. S9(a)[31].

**Acknowledgements**

Imaging work at the ALS was supported by the U.S. Department of Energy (DE-AC02-05CH11231). J.A.B. and E.E.F. acknowledge support for sample fabrication, modeling, testing and synchrotron measurements by the Quantum Materials for Energy Efficient Neuromorphic-Computing Energy Frontier Research Center funded by DOE, Office of Science, BES under Award No. DE-SC0019273. S. A. M. acknowledges support by the U. S. Office of Naval Research, In-House Laboratory Independent Research Program. M.-Y.I. acknowledges support by Lawrence Berkeley National Laboratory through the Laboratory Directed Research and Development (LDRD) Program and by the National Research Foundation of Korea (NRF) grant funded by the Korea government (MSIT)(NRF-2019R1A2C2002996, NRF-2016M3D1A1027831, and NRF-2019K1A3A7A09033400). This work was performed in part at the San Diego Nanotechnology

Infrastructure (SDNI), a member of the National Nanotechnology Coordinated Infrastructure, which is supported by the National Science Foundation (ECCS-1542148).

**Main text figures:**

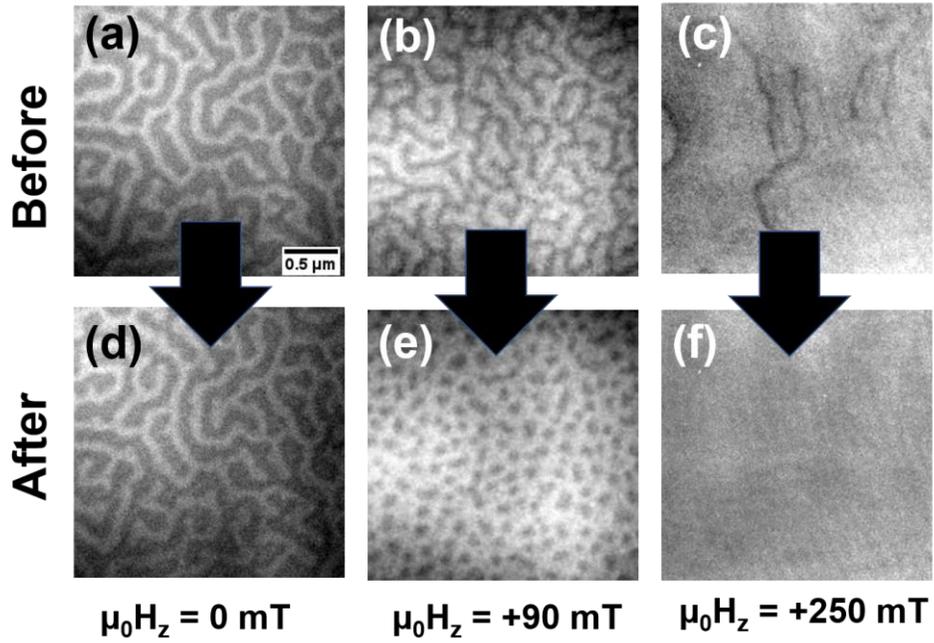

Figure 1: MTXM images illustrating the effect of a 60-μs-long current pulse of density $1.7 \times 10^{10}$ A/m$^2$ on the domain morphology of a [Co (0.7 nm)/ Ni (0.5 nm)/ Pt (0.7 nm)]$_{20}$ wire for $\mu_0 H_z = 0$ (a, d), +50 (b, e), and +250 mT (c, f) before and after the current is applied. The indicated perpendicular field $\mu_0 H_z$ was maintained before, during, and after the current pulse was applied. Low-contrast vertical lines observed in the foreground of (c, f) are artifacts of the MTXM optics.

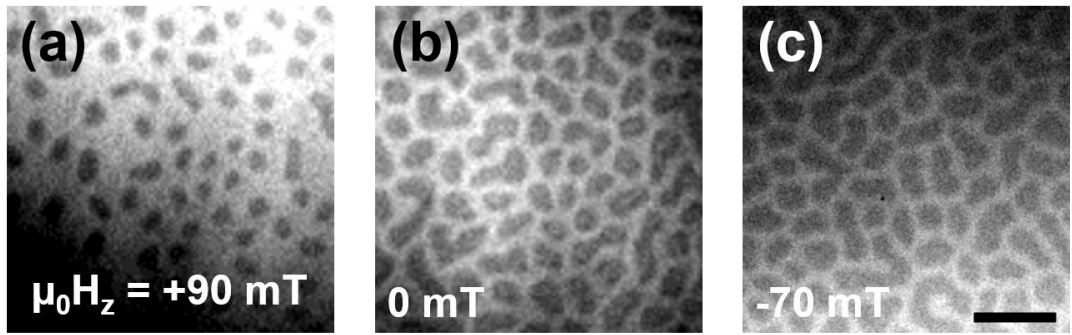

Figure 2: MTXM images of the skyrmion phase in the [Co (0.7 nm)/ Ni (0.5 nm)/ Pt (0.7 nm)]$_{20}$ sample as the out-of-plane field strength was varied. Before imaging, the stripe-to-skyrmion transformation was accomplished by applying a 60-μs current pulse of density 1.7 x 10$^{10}$ A/m$^2$ in a field of $\mu_0 H_z$ = +190 mT. (bar = 500 nm)

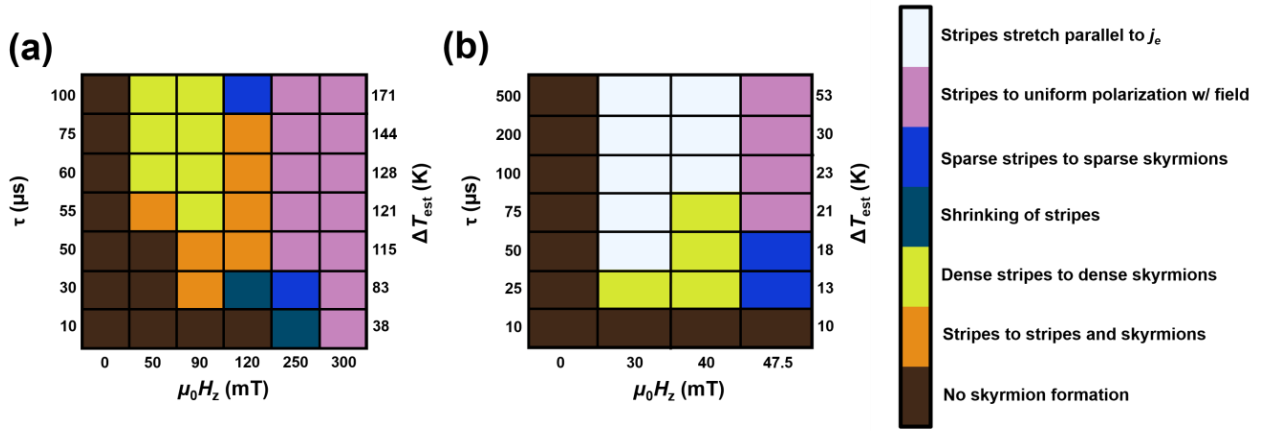

Figure 3: $\tau$-$\mu_0 H_z$ phase maps for the (a) [Co (0.7 nm)/ Ni (0.5 nm)/ Pt (0.7 nm)]$_{20}$ and (b) [Co (0.7 nm)/ Tb (0.4 nm)/ Ni (0.5 nm)/ Pt (0.7 nm)]$_{20}$ samples indicating the change in morphology observed after pulses of $j = 1.7 \times 10^{10}$ A/m$^2$ and $j = 2.4 \times 10^9$ A/m$^9$ (respective to each sample) were applied for various durations ($\tau$) and out-of-plane magnetic field strengths ($\mu_0 H_z$). The experimentally estimated temperature increase $\Delta T_{est}$ for each pulse duration is also shown (see Methods, Supp. Fig. S2[31]). Before each measurement, the magnetic field was swept to both positive and negative saturation before setting the target magnetic field strength for the trial. Note that both the $\tau$ and $\mu_0 H_z$ -axes are not linearly scaled.

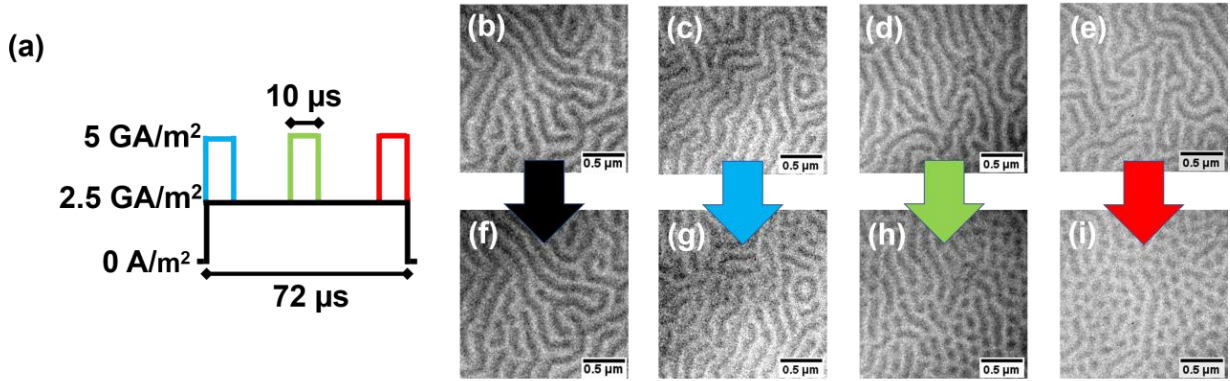

Figure 4: (a) Depiction of the composite pulse employed to delineate between Joule heating and spin-orbit torques. (b-i) MTXM images of a [Co (0.4 nm)/ Tb (0.4 nm)/ Co (0.4 nm)/ Ni (0.5 nm)/ Pt (0.7 nm)]-type sample illustrating the effects delivering the same Joule heating and same net electron flow by changing the relative delay of the spike pulse to the (c, g) beginning, (d, h) middle, and (e, i) end of the baseline pulse in a field of $\mu_0 H_z = +173$ mT. (b, f) demonstrates the effect of only applying the baseline pulse.

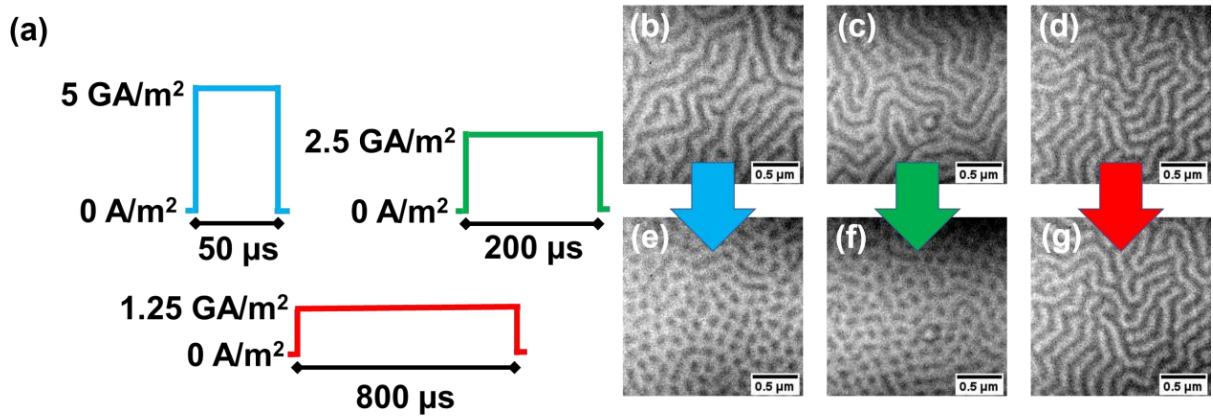

Figure 5: (a) Schematic depictions of the variation in pulse duration τ and current density $j$ used to examine the impact of supplying an equivalent Joule heating using 50 µs (blue), 200 µs (green), and 800 µs (red)-long pulses. (b-g) MTXM images of a [Co (0.4 nm)/ Tb (0.4 nm)/ Co (0.4 nm)/ Ni (0.5 nm)/ Pt (0.7 nm)]-type sample illustrating the effect of delivering the 50 µs (b,e), 200 µs (c,f), and 800 µs (d,g)-long pulses illustrated in (a) in a perpendicular field of $\mu_0 H_z = +173$ mT.

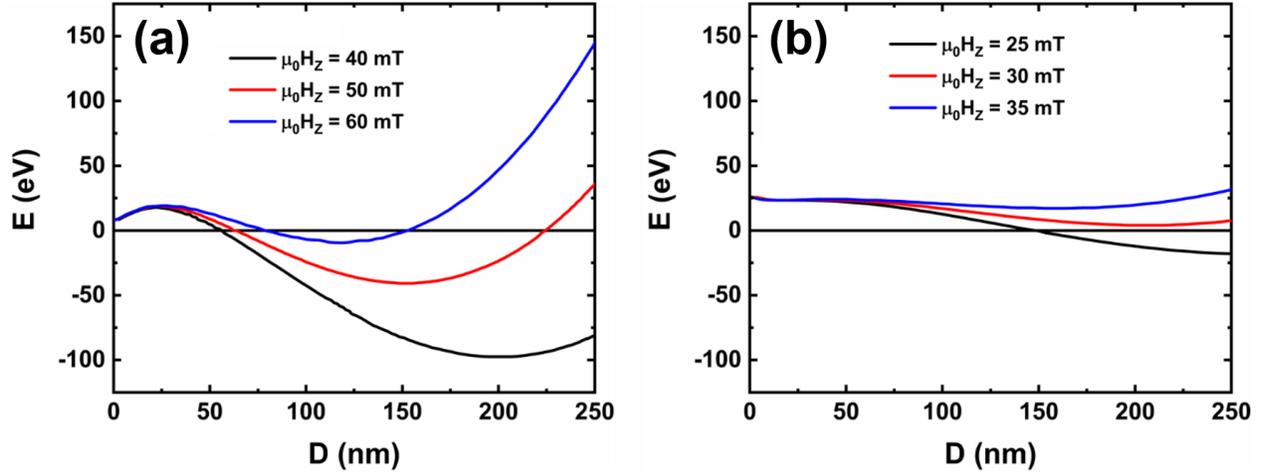

Figure 6: The energy profile of isolated bubble domains at several perpendicular field strengths $\mu_0 H_z$ in proximity to the lowest $\mu_0 H_z$ necessary for skyrmion nucleation in the (a) [Co (0.7)/ Ni (0.5)/Pt (0.7)]$_{20}$ and the (b) [Co (0.7)/ Ni (0.5)/Pt (0.7)]$_{20}$ samples (thicknesses in nm) . The energy profile was calculated using the analytical treatment proposed in Ref 36. In calculating (a), sample properties of $M_S$ = 1000 emu/cc, $K_u$ = 1.1 x 10$^7$ J/m$^3$, DMI energy density of 0.6 mJ/m$^2$, and exchange stiffness $A$ = 10 pJ/m were used. In calculating (b), sample properties of $M_S$ = 500 emu/cc, $K_u$ = 1.7 x 10$^6$ J/m$^3$, DMI energy density of 0.6 mJ/m$^2$, and exchange stiffness $A$ = 10 pJ/m were used.

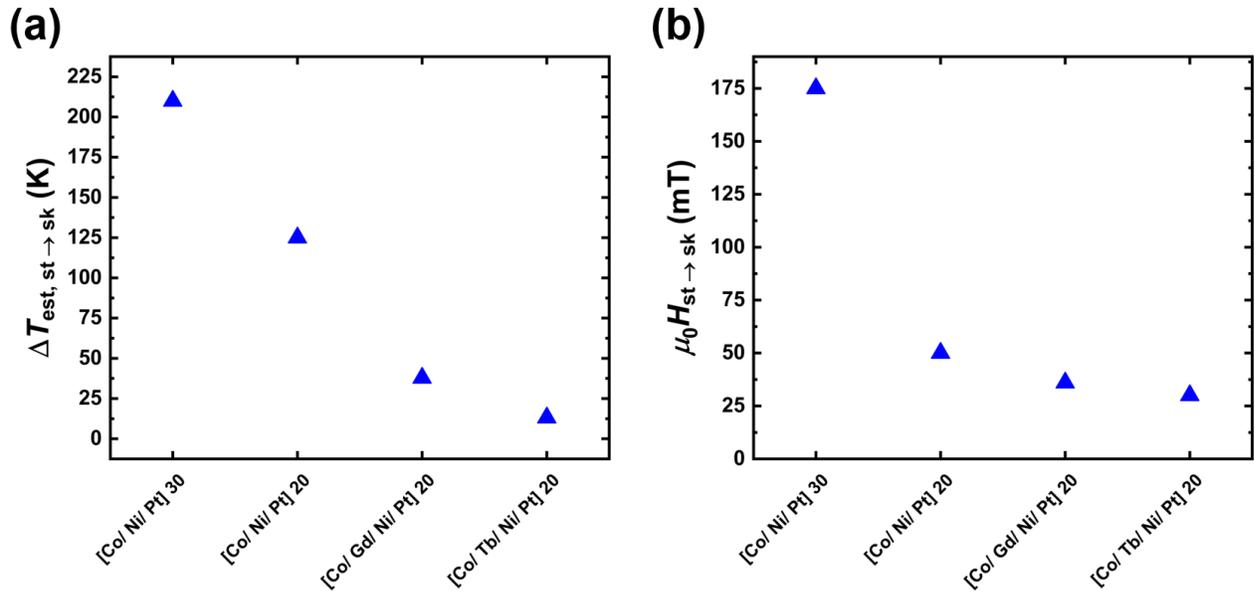

Figure 7: (a) The experimentally estimated temperature change due to Joule heating ($\Delta T_{\text{est, st} \to \text{sk}}$) and (b) perpendicular field ($\mu_0 H_{\text{st} \to \text{sk}}$) necessary to initiate a complete stripe-to-skyrmion transformation in the [Co/ Ni/ Pt]-based samples reported here. Data corresponds to the necessary electrical pulse characteristics at the lowest possible magnetic field.

**Supplemental Figures:**

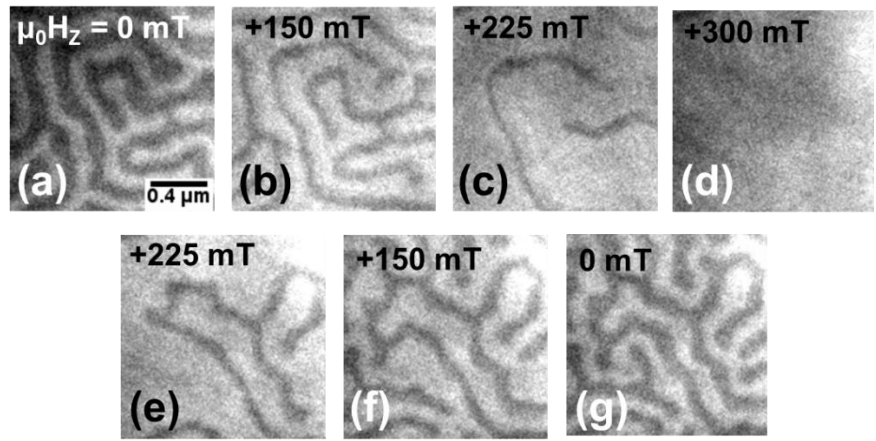

Supplemental Figure S1: Magnetic transmission X-ray microscopy (MTXM) images of the magnetic domain morphology as an out-of-plane magnetic field ($\mu_0 H_Z$) was swept from zero towards positive saturation then back to zero in a Ta (2)/ Pt (5)/[Co (0.7)/ Ni (0.5)/Pt (0.7)]$_{20}$/ Ta (5) (thicknesses in nm) sample.

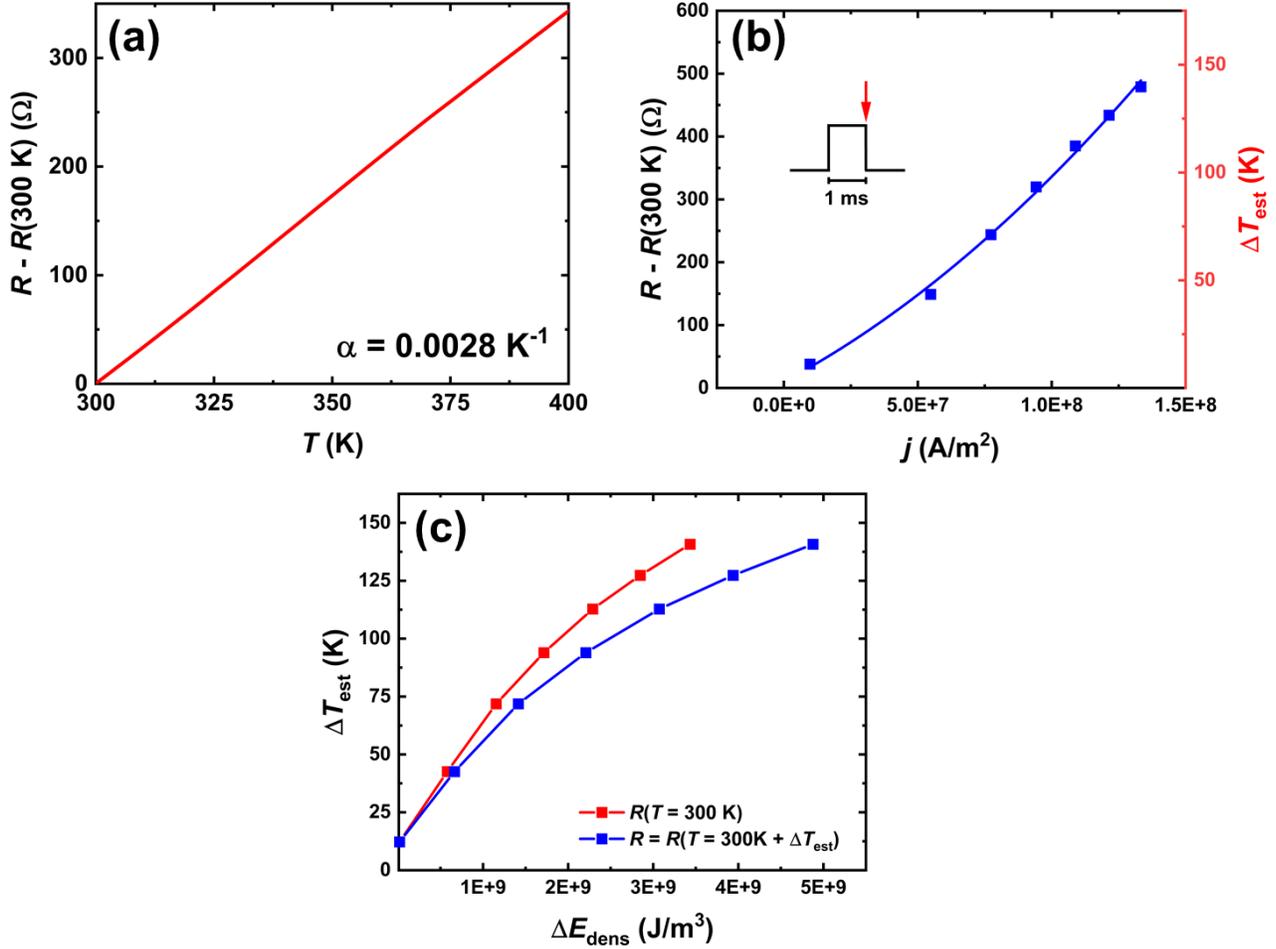

Supplemental Figure S2: (a) Change in resistance (relative to the resistance at 300 K) as a function of temperature for a Ta (2)/ Pt (5)/[Co (0.7)/ Ni (0.5)/Pt (0.7)]$_{20}$/ Ta (5) device patterned in to a 3 mm-long, 10 µm-wide wire on a SiN membrane used to calculate the temperature coefficient of resistance $\alpha$. The data in (a) was collected using an AC lock-in technique, with an excitation current amplitude of 0.1 mA applied at a frequency of 57.98 Hz. (b) Change in resistance (relative to resistance at 300 K) and estimated increase in temperature $\Delta T_{est}$ due to Joule heating for several current densities for the sample as in (a). For each current density in (b), the current was applied for 1 ms before a four-point resistance measurement was made (as indicated by the red arrow in the inset) before the current was turned off. (c) $\Delta T_{est}$ as a function of Joule heating energy density $\Delta E_{dens}$ for the same sample as (a-b). To place an upper and lower limit on the amount of Joule heating for a particular current density, $\Delta E_{dens}$ was calculated assuming both the resistance of the sample at room temperature, as well as the resistance at the elevated temperature estimated from (b).

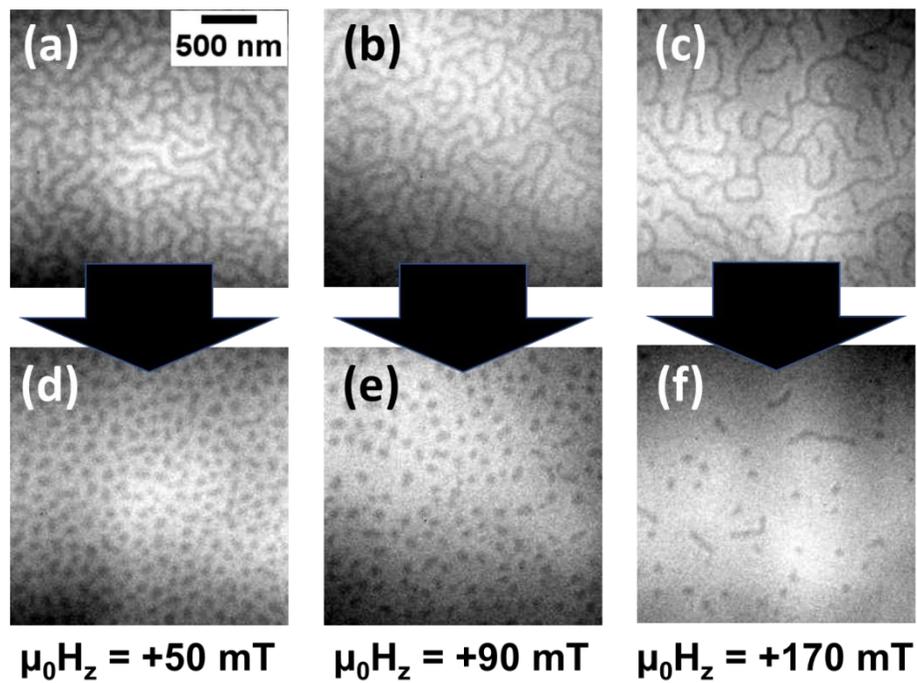

Supplemental Figure S3: MTXM images illustrating the effect of applying a $\tau = 100$ µs-long current pulse with density $1.7 \times 10^{10}$ A/m$^2$ on the domain morphology of a Ta (2)/ Pt (5)/[Co (0.7)/ Ni (0.5)/Pt (0.7)]$_{20}$/ Ta (5) (thicknesses in nm) sample patterned into a wire for applied magnetic fields of $\mu_0 H_z = +50$ (a, d), +90 (b, e), and +120 mT (c, f).

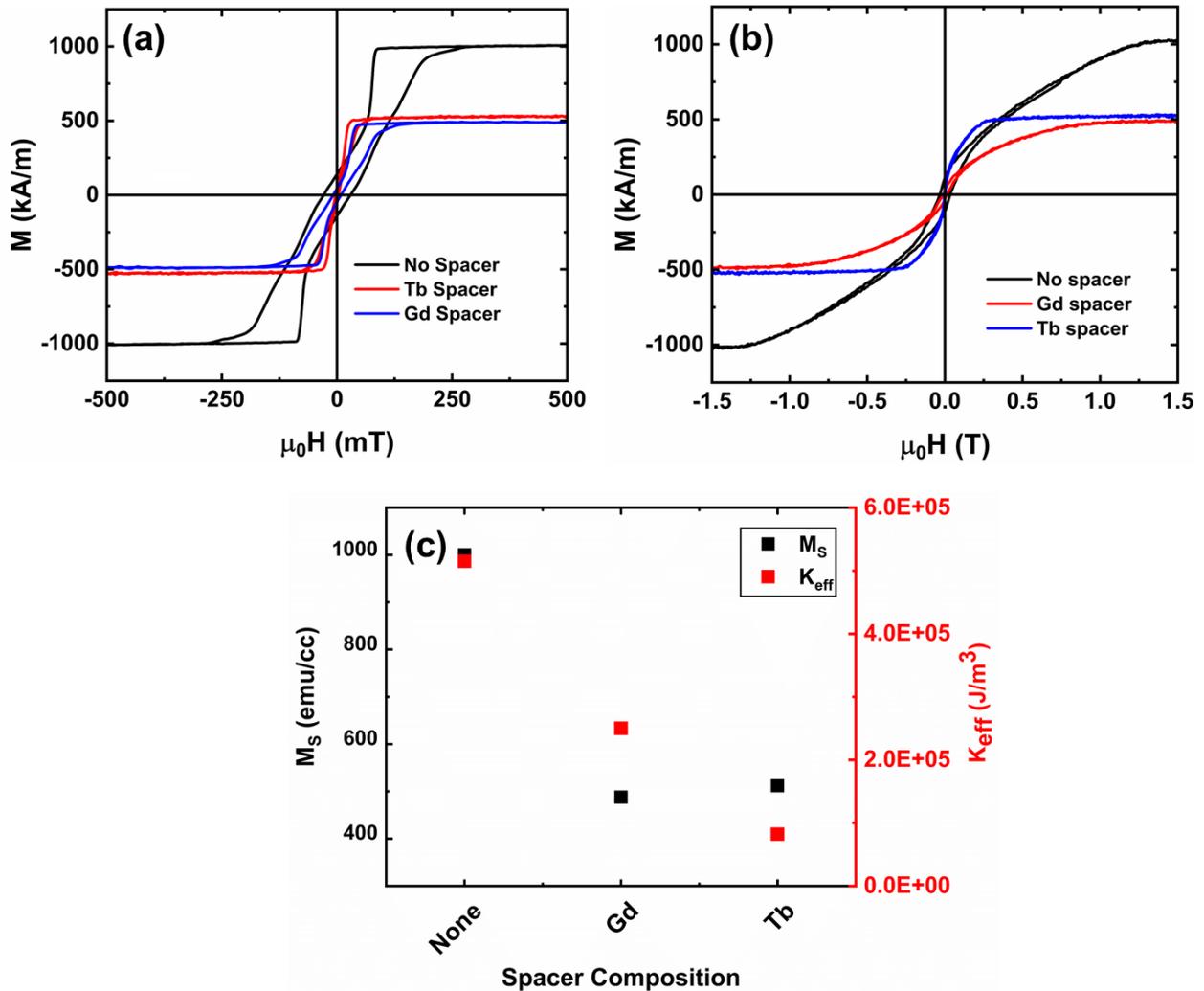

Supplemental Figure S4: (a) Out-of-plane and (b) in-plane vibrating sample magnetometry characterization of the room temperature magnetic properties of the Ta (2)/ Pt (5)/[Co (0.7)/Spacer (0.4)/Ni (0.5)/Pt (0.7)]$_{20}$/ Ta (5) (thicknesses in nm) samples. (c) Saturation magnetization ($M_S$) and effective perpendicular anisotropy energy density ($K_{eff}$) of the samples determined from (a) and (b).

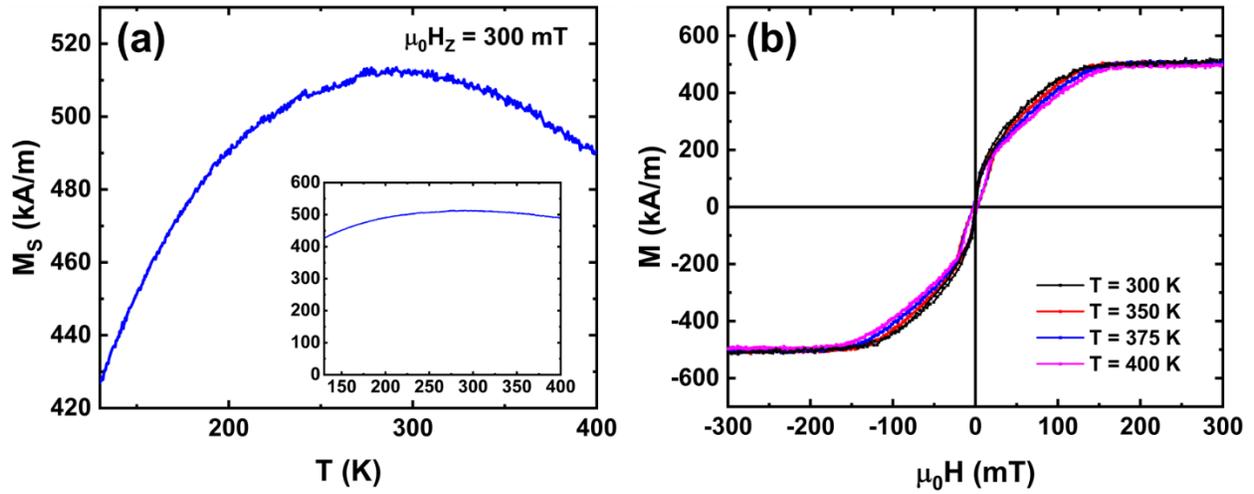

Supplemental Figure S5: (a) The saturation magnetization $M_S$ as a function of temperature, and (b) in-plane hysteresis loops at selected temperatures of an unpatterned Ta (2)/ Pt (5)/[Co (0.7)/Tb (0.4)/Ni (0.5)/Pt (0.7)]$_{20}$/ Ta (5) (thicknesses in nm) sample, determined using vibrating sample magnetometry.

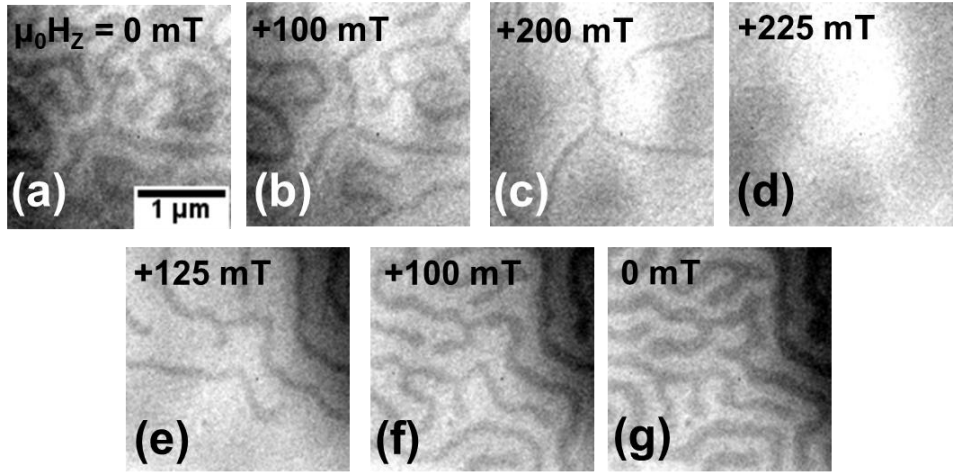

Supplemental Figure S6: MTXM images of the magnetic domain morphology as an out-of-plane magnetic field ($\mu_0 H_Z$) was swept from zero towards positive saturation then back to zero in the Ta (2)/ Pt (5)/[Co (0.7)/Tb (0.4)/Ni (0.5)/Pt (0.7)]$_{20}$/ Ta (5) (thicknesses in nm) sample.

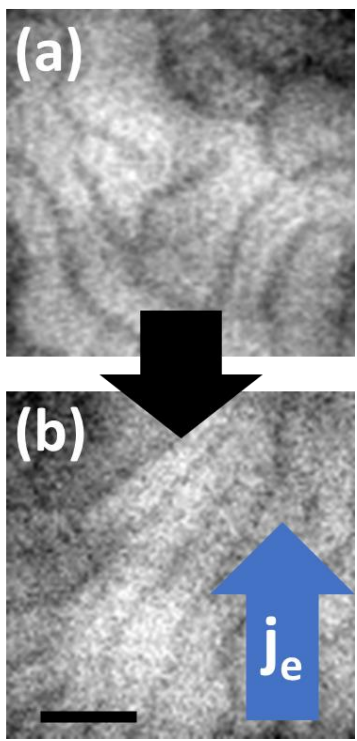

Supplemental Figure S7: MTXM images depicting the change in domain morphology of the Ta (2)/ Pt (5)/[Co (0.7)/Tb (0.4)/Ni (0.5)/Pt (0.7)]$_{20}$/ Ta (5) (thicknesses in nm) sample before (a) and after (b) a 200-µs pulse of current density 2.4 x 10$^9$ A/m$^2$ was applied in an applied magnetic field of + 40 mT. The blue arrow indicates the electron flow direction $j_e$. (bar = 500 nm)

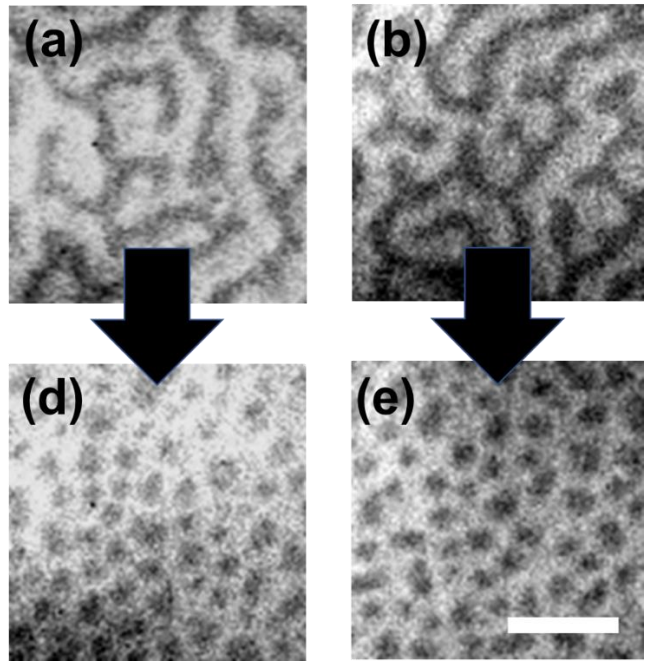

Supplemental Figure S8: MTXM images illustrating the effect of applying a $\tau = 50$ µs-long current pulse with density $1.7 \times 10^{10}$ A/m$^2$ on the domain morphology of the (a) Ta (2)/Pt (5)/[Co (0.7)/ Ni (0.5)/Pt (0.7)]$_{20}$/Ta (5) sample and the (b) Ta (2)/ Pt (5)/[Co (0.7)/ Ni (0.5)/Pt (0.7)]$_{20}$/ Pt (5) (thicknesses in nm) samples in an applied magnetic field of +50 mT. (scale bar = 400 nm)

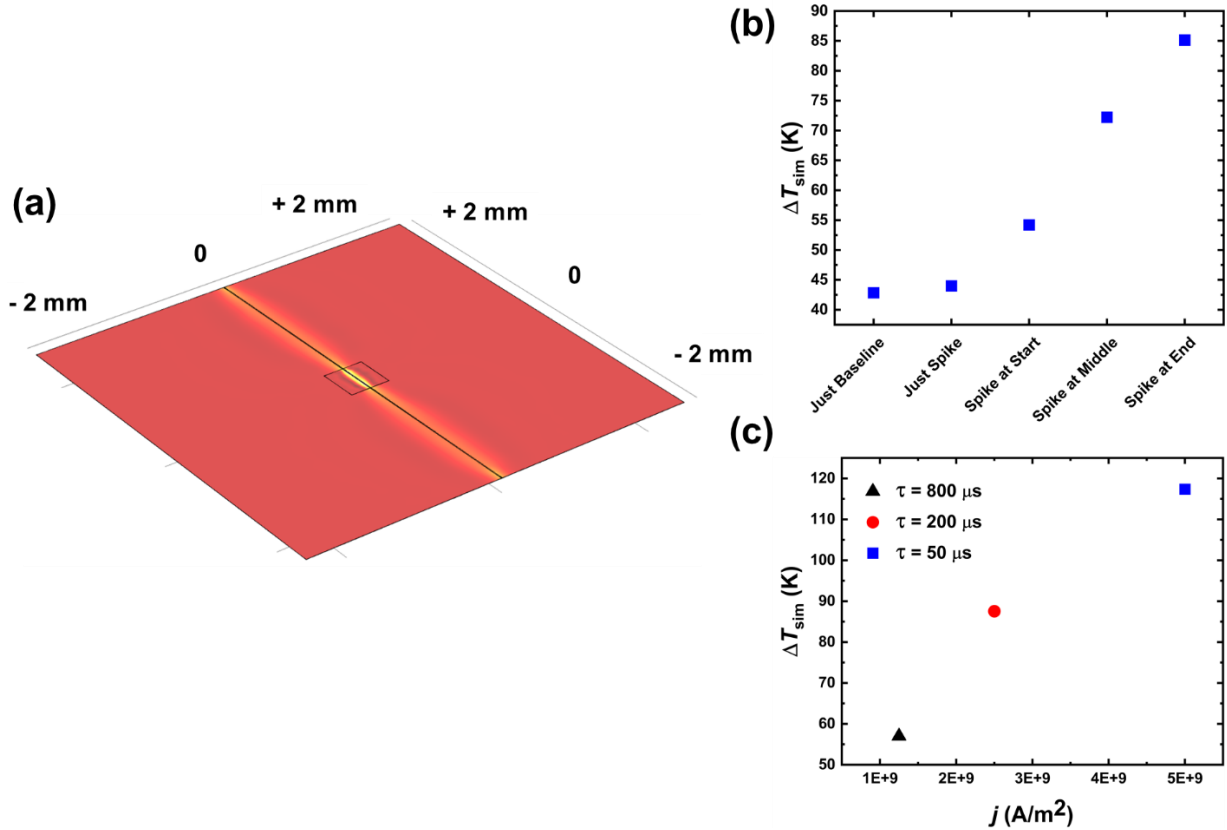

Supplemental Figure S9: (a) Experimental geometry employed in COMSOL Multiphysics simulations of the temperature change due to Joule heating in [Co/Ni/Pt]-based multilayers. The model features a 10 µm x 1 mm x 50 nm Pt wire with a resistivity value that approximates a Ta(2)/ Pt (5)/[Co(0.7 nm)/ Ni (0.5 nm)/ Pt (0.7)]$_{20}$/ Ta (5) (thicknesses in nm) multilayer placed on top of a 0.5 mm x 0.5 mm x 100 nm Si$_3$N$_4$ window, surrounded by a 5 mm x 5 mm x 0.5 mm SiOx frame. Convective cooling to ambient ($K = 5$ Wm$^{-2}$K$^{-1}$) was modeled at all structure boundaries. (b) Simulated temperature changes $\Delta T_{sim}$ in response to the composite pulse configurations indicated in Fig. 4(a) of the main text, as well as $\Delta T_{est}$ from applying only the baseline pulse or spike pulse. (c) $\Delta T_{sim}$ in response to the pulses of varying duration $\tau$ and current density $j$ but equivalent Joule heating depicted in Fig. 5(a). The initial temperature of the sample was set to 300 K for all simulations.

simulations of the differences in temperature changes when using electrical pulses of differing current density and duration.